\documentclass[preprint,12pt]{elsarticle}

\usepackage{hyperref}

\journal{N.I.M. A}\pdfoutput=1
\usepackage{graphicx}
\usepackage{color}
\usepackage{mathtools}
\usepackage[left]{lineno}
\usepackage{hyperref}

\begin{document} 
\begin{frontmatter}

\title{$^9$Li and $^8$He decays in GEANT4}
\def\A{\kern+.6ex\lower.42ex\hbox{$\scriptstyle \iota$}\kern-1.20ex a}
\def\E{\kern+.5ex\lower.42ex\hbox{$\scriptstyle \iota$}\kern-1.10ex e}
\author[cenbg]{C.Jollet}
\author[cenbg]{A.Meregaglia}

\address[cenbg]{CENBG, Universit\'{e} de Bordeaux, CNRS/IN2P3, 33175 Gradignan, France}


\begin{abstract}
The decays of cosmogenic nuclei such as $^9$Li and $^8$He represent one of the largest irreducible backgrounds for reactor antineutrino experiments. The correct treatment of such decays are of fundamental importance in the study of cosmogenic backgrounds and in their rejection, hence the full chain of intermediate excited states must be accounted for.  Currently the treatment in GEANT4 of the modelling of de-excitation of $^9$Be and $^8$Li, which are the daughter nuclei of $^9$Li and $^8$He respectively, is not correct. $^9$Be excited states should break into a neutron and two $\alpha$'s, and $^8$Li excited states should emit a neutron and possibly an $\alpha$ and a triton depending on the decay chain, whereas in GEANT4 they both reach the ground state by emitting a gamma. Based on the available nuclear measurements we included the correct treatment of $^9$Li and $^8$He decays in GEANT4 and compared the obtained results with the spectra published by the Double Chooz collaboration finding an excellent agreement.
\end{abstract}

\begin{keyword}
neutrino \sep reactor \sep cosmogenic background \sep GEANT4


\end{keyword}


\end{frontmatter}

\section{Introduction}
The cosmogenic production of $^9$Li and $^8$He and their following decays represent one of the largest irreducible backgrounds for reactor antineutrino experiments such as Double Chooz~\cite{Abe:2011fz} or JUNO~\cite{An:2015jdp}. Such experiments aim at the neutrino detection through the so called inverse beta decay process (IBD) i.e. $\bar{\nu}_e + p \to e^+ + n$, observing the time correlation between the prompt signal given by the positron and the delayed one given by the neutron capture on hydrogen or gadolinium.
Several possible decay schemes of  $^9$Li and $^8$He have an electron and a neutron in the final state which mimic exactly the expected signal from the IBD. Most of the time an $\alpha$ particle is also emitted which could be used to reject such a background, however due to the large quenching it is often difficult to identify its energy deposition contribution which is hidden by the electron one, reducing the effectiveness of this additional handle.\\
To study such a background and develop appropriate tools to reject it, a complete simulation of the possible decay schemes including the correct branching ratios (BR) is fundamental. Such a treatment is today missing in standard simulation tools such as GEANT4~\cite{Agostinelli:2002hh}. In this work we describe in detail the different decay schemes, their implementation in GEANT4, and we present a validation of the Monte-Carlo comparing the results to published Double Chooz data.

\section{Current status}

In the latest release of the GEANT4 software i.e. GEANT4 10.5p01 cosmogenic nuclei undergo $\beta$ decay reaching the correct  state i.e. $^9$Li $\to$ $^9$Be + $e^-$ + $\bar{\nu}_e$ and  $^8$He $\to$ $^8$Li + $e^-$ + $\bar{\nu}_e$. However, when the $^9$Be and $^8$Li nuclei are produced in an excited state, they undergo de-excitation to the ground state through gamma emission and the correct decay chains including $\alpha$'s and neutrons are not accounted for. The decay chains are driven by the radioactive decay data used in the simulation which are coded in RadioactiveDecay data version 5.3~\cite{RD}.  Of course this treatment is not appropriate for antineutrino reactor experiments since they never result in background (electron plus neutron emission) contrarily to real life.\\
Experiments found ad-hoc workarounds to simulate cosmogenic background which however required a lot of specific developments from which the community can not benefit being private tools. In Double Chooz the solution which was adopted is to have an external generator which produces directly the output of the decays namely $\alpha$, electrons and neutrons, taking into account the large uncertainties on the energy levels and branching ratios. With such a generator a good agreement between data and Monte-Carlo was obtained as published in Ref.~\cite{deKerret:2018fqd}. Unfortunately the portability of the generator to other experiments is not straightforward including the fact that it is intellectual property of the collaboration. 

\section{Decay schemes}

We decided to gather all the available existing knowledge in terms of decay schemes and branching ratios and to include them in GEANT4 for an easy and public access.\\
Despite the large number of experiments which studied the behaviour of $^9$Li and $^8$He the actual knowledge is not complete for some excited levels in the decay chain. We were therefore forced to make some assumption based on the existing publications which might not be fully correct. In the following we will clearly state what we know from experiments and what was reasonably assumed.

\subsection{$^9$Li}

\begin{figure} [tp]
\begin{center}
\includegraphics[width=0.75\textwidth]{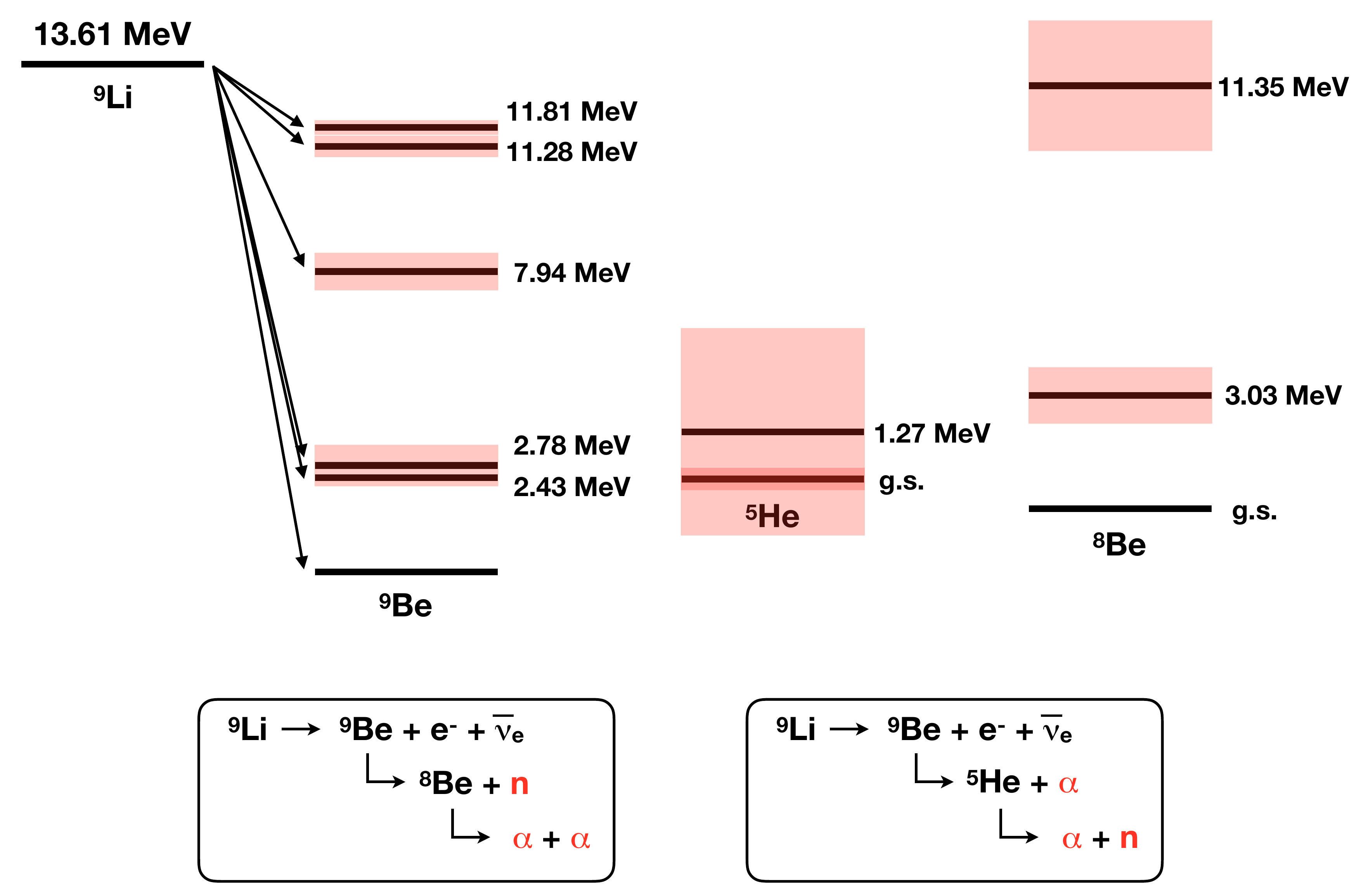}
\caption{{\it Decay scheme of $^9$Li including the states of $^8$Be and $^5$He. The shaded bands represent the width of the excited states whereas the decay chains going through the $^9$Be excited states are shown in the two boxes.}}
\label{fig:1}
\end{center}
\end{figure}

The mass difference between $^9$Li and $^9$Be is 13.61 MeV as can be seen in Fig.~\ref{fig:1}.  In 49.2\% of the cases $^9$Li decays directly to the $^9$Be ground state, not representing a source of background for antineutrino reactor experiments. The remaining 50.8\% go through an excited state of $^9$Be which has an energy larger than 1.57 MeV and which undergoes a decay through $\alpha + \alpha + n$ (1.57 MeV is energy difference between $\alpha + \alpha + n$ and the $^9$Be ground state).\\
All the excited $^9$Be states decay through an $\alpha + \alpha + n$ in two ways: the first possibility is a neutron emission going to a $^8$Be state (excited or ground state) which in turn decays into two $\alpha$'s. The second possibility is an $\alpha$ decay to $^5$He which in turn decays into $\alpha + n$. A schematic view is shown in Fig.~\ref{fig:1}.\\  
The different branching ratios and relative errors have been measured over the years by several experiments. In the following, the excitation levels of the different nuclei are identified by their energy in MeV given in parenthesis or by the label ``g.s.'' for ground state. In Tab.~\ref{tab:1} we quote the most up to date decay modes and the relative references whereas the level of daughter nuclei can be found in Tab.~\ref{tab:1b}. For each state the associated error is quoted: in can be seen that for some states such as $^5$He (1.27) or $^8$Be (11.35) the uncertainties are quite large.\\
 It must be noted that for the excited $^9$Be state at 7.94~MeV, Ref.~\cite{Prezado:2005odo} states: ``The 7.94 MeV level has been found to decay mainly through $^5$He (g.s.) with 10\% contribution to $^8$Be (g.s.). Contributions from $^8$Be (3.03) are found to be small ($< 20$\%)'' and the uncertainties are not quoted. We assumed that ``mainly'' means 80\% and the small contribution to $^8$Be (3.03) is 10\%. Similarly, for the excited $^9$Be state at 2.78~MeV, Ref.~\cite{Prezado:2005odo} states that the feeding of the $^8$Be (g.s.) has not been measured due to energy threshold limitation and that ``Contributions from the $^8$Be (3.03) and the $^5$He (g.s.) channels are taken as 0.75 and 0.25, respectively'' without giving any indication of the feeding of the $^5$He (1.27) state and on the BR uncertainty. We therefore neglected the $^5$He (1.27) contribution and assumed a probability of 60\% for $^8$Be (3.03) and 15\% for $^8$Be (g.s.). Such assumption is motivated by the fact that, although in Ref.~\cite{Prezado:2005odo} the decay to the ground state of $^8$Be was not observed, this is probably due to the detection energy threshold whereas such a transition had been previously observed~\cite{Cocke:1968jsh}.

\begin{table}[htp]
\begin{center}
\begin{tabular}{|c|c|c|c|c|c|}
\hline
$^9$Be state &  $\Delta$E & $\beta$ decay & Decay mode & Decay mode  & Reference\\
(MeV) & (MeV)  & BR &  & BR & \\

\hline
& &  & $^5$He (g.s.) + $\alpha$ & 28\% $\pm$ 6\% & \\
& &  & $^5$He (1.27) + $\alpha$ & 47\% $\pm$ 6\% & \\
11.81  & 0.4 & 2.7\% & $^8$Be (g.s.) + n & 2\% $\pm$ 1\% & \cite{Tilley:2004zz,Prezado:2003mqr}\\
& &  & $^8$Be (3.03) + n  & 11\% $\pm$ 6\% & \\
& &  & $^8$Be (11.35) + n  & 12\% $\pm$ 8\% & \\
\hline
& &  & $^5$He (g.s.) + $\alpha$ & 76\% $\pm$ 30\% & \\
11.28  &0.58& 1.1\% & $^8$Be (g.s.) + n & 3\% $\pm$ 1\% & \cite{Tilley:2004zz,Brown:2007zze}\\
& &  & $^8$Be (3.03) + n  & 21\%$\pm$ 8\% & \\
\hline
& &  & $^5$He (g.s.) + $\alpha$ & 80\% & \cite{Prezado:2005odo,Tilley:2004zz} \\
7.94  &1.& 1.5\% & $^8$Be (g.s.) + n & 10\%  & + our \\
& &  & $^8$Be (3.03) + n  & 10\% &assumptions\\
\hline
 &&  & $^5$He (g.s.) + $\alpha$ & 25\% & \cite{Prezado:2005odo,Tilley:2004zz}\\
2.78  &1.1& 15.8\% & $^8$Be (g.s.) + n & 15\%  & + our \\
& &  & $^8$Be (3.03) + n  & 60\% & assumptions\\
\hline
& &  & $^5$He (g.s.) + $\alpha$ & 2.5\% $\pm$ 2.5\% &\cite{Tilley:2004zz,Papka:2007sa} \\
2.43  &$<$0.01& 29.7\% & $^8$Be (g.s.) + n & 11\%. $\pm$ 2\% & + our\\
 &&  & $^8$Be (3.03) + n  & 86.5\%$\pm$ 4.5\% & assumptions\\
\hline
\end{tabular}
\end{center}
\caption{{\it Branching ratios and relative errors for the different decay channels of $^9$Li. In the last column the references are quoted.}}
\label{tab:1}
\end{table}%

\begin{table}[htp]
\begin{center}
\begin{tabular}{|c|c|c|c|}
\hline
Nucleus & State &  $\Delta$E &  Reference\\
 & (MeV)  &(MeV)  &   \\

\hline
 $^5$He & g.s. & 0.6 &  \cite{Tilley:2004zz,Prezado:2003mqr}  \\
 $^5$He & 1.27 & 5.57 &  \cite{Tilley:2004zz,Prezado:2003mqr,Tilley:2002vg}  \\
 $^8$Be & 3.03 & 1.51 &  \cite{Tilley:2004zz,Prezado:2003mqr}  \\
 $^8$Be & 11.35 & 3.5 &  \cite{Tilley:2004zz,Prezado:2003mqr}  \\
\hline
\end{tabular}
\end{center}
\caption{{\it States with their excitation energy and uncertainty of nuclei of interest for the $^9$Li decay.}}
\label{tab:1b}
\end{table}%

\subsection{$^8$He}

\begin{figure} [tp]
\begin{center}
\includegraphics[width=0.75\textwidth]{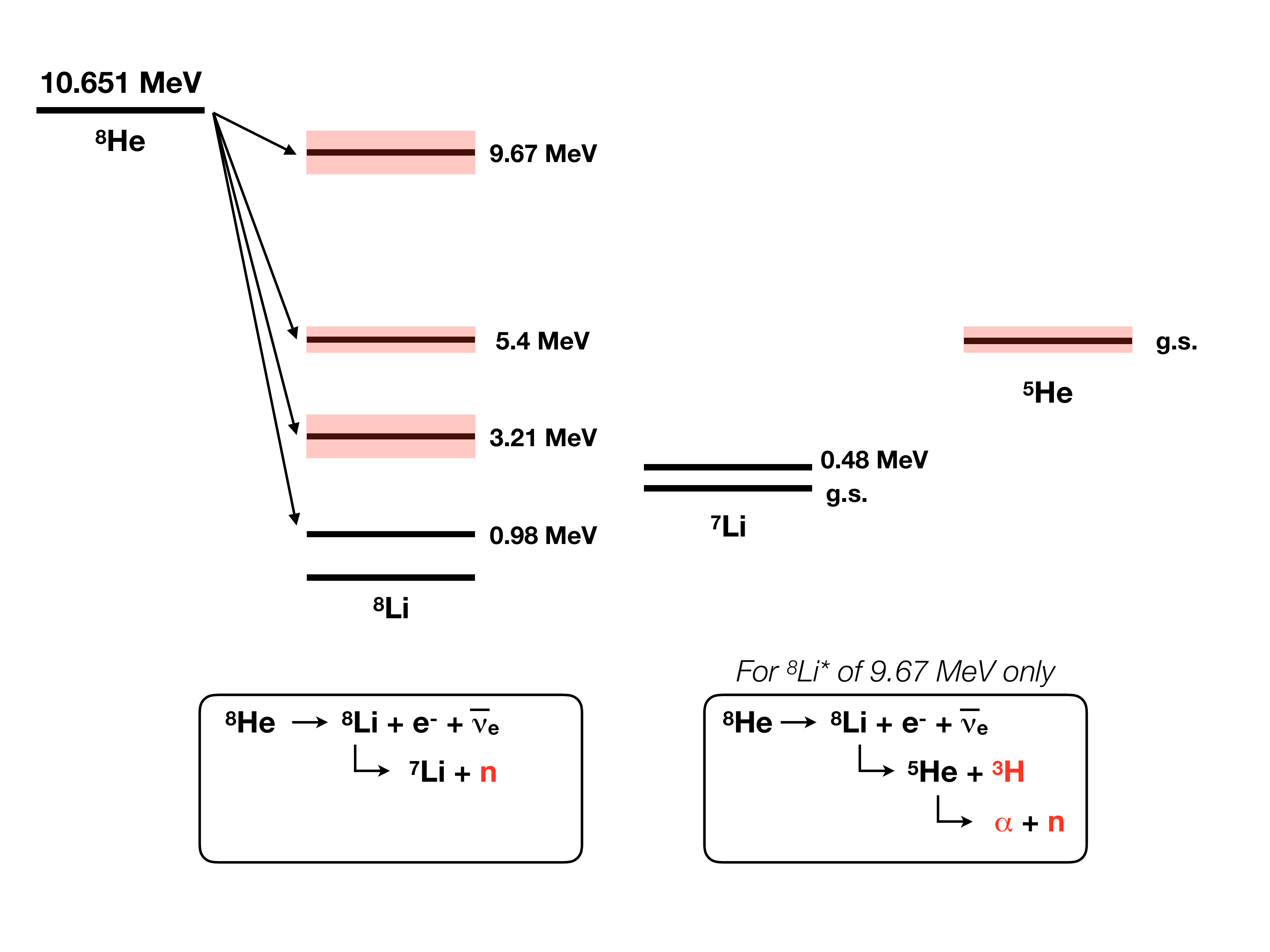}
\caption{{\it Decay scheme of $^8$He including the states of $^7$Li and $^5$He. The shaded bands represent the width of the excited states whereas the decay chains going through $^8$Li states with energy larger than 0.98 MeV are shown in the two boxes.}}
\label{fig:2}
\end{center}
\end{figure}

The produced $^8$He undergoes $\beta$ decay reaching an excited state of $^8$Li. Only the state with the highest excitation energy of 9.67~MeV could decay into triton and $^5$He which in turn decays into $\alpha$ and neutron. All the other states, apart from the excited one at 0.98~MeV, will decay via neutron emission to $^7$Li. A schematic view is shown in Fig.~\ref{fig:2}: as it was the case for $^9$Li the final products include an electron and a neutron mimicking the signal signature for antineutrino reactor experiments.\\ 
The different branching ratios and relative errors have been measured over the years by several experiments although the uncertainties are much larger with respect to the ones related to $^9$Li decays. In Tab.~\ref{tab:2} we quote the most up to date values and the relative references whereas the level of daughter nuclei can be found in Tab.~\ref{tab:2b}. In the case of $^8$He there are fewer measurements available with respect to $^9$Li, and the uncertainty on the energy levels is higher. Indeed the states of 3.21~MeV and 5.4~MeV are sometimes quoted as states at 3.08~MeV and 5.15~MeV respectively~\cite{Tilley:2004zz,Barker:1988biv}. 
The BR of $^8$He decays into $^8$Li at energies of 5.4~MeV and 3.21~MeV was measured to 16\% in total but no measurement exists for the individual states. We assumed an equal probability and therefore a BR of 8\%. For what concerns the decay of $^8$Li to $^7$Li there are no measurements yielding the BR to the ground state and to the excited level at 0.48~MeV: we therefore assumed an equal probability for the BR of the two channels. In any case the difference is the emission of a gamma of 0.48~MeV which has a negligible impact on the overall obtained energy spectrum. 

\begin{table}[htp]
\begin{center}
\begin{tabular}{|c|c|c|c|c|c|}
\hline
$^8$Li state &  $\Delta$E & $\beta$ decay & Decay mode & Decay mode  & Reference\\
(MeV) & (MeV)  & BR &  & BR & \\
\hline
& &  & $^3$H +$^5$He (g.s.) & 80\% & \\
9.67 & 1 & 0.9\% & $^7$Li (0.48) + n & 10\%  & \cite{Tilley:2004zz,Borge:1993st}\\
& &  & $^7$Li (g.s.) + n  & 10\% & + our assumptions\\
\hline
5.4  &0.65 & 8.0\% & $^7$Li (0.48) + n & 50\%  & \cite{Tilley:2004zz,Borge:1993st}\\
 &&  & $^7$Li (g.s.) + n  & 50\%  & + our assumptions\\
\hline
3.21  &1& 8.0\% & $^7$Li (0.48) + n & 50\% & \cite{Tilley:2004zz,Borge:1993st}\\
 & & & $^7$Li (g.s.) + n  & 50\%  &+ our assumptions\\
\hline
\end{tabular}
\end{center}
\caption{{\it Branching ratios and relative errors for the different decay channels of $^8$He. In the last column the references are quoted.}}
\label{tab:2}
\end{table}%

\begin{table}[htp]
\begin{center}
\begin{tabular}{|c|c|c|c|}
\hline
Nucleus & State &  $\Delta$E &  Reference\\
 & (MeV)  &(MeV)  &   \\

\hline
 $^5$He & g.s. & 0.6 &  \cite{Tilley:2004zz,Prezado:2003mqr}  \\
 $^7$Li & g.s. & $<$0.01 &  \cite{Tilley:2004zz,Prezado:2003mqr}  \\
 $^7$Li & 0.48 & $<$0.01 &  \cite{Tilley:2004zz,Prezado:2003mqr}  \\
\hline
\end{tabular}
\end{center}
\caption{{\it States with their excitation energy and uncertainty of nuclei of interest for the $^8$He decay.}}
\label{tab:2b}
\end{table}%

\section{GEANT4 implementation and data comparison}
Once the different branching ratios and decay schemes have been identified, the most effective implementation in GEANT4 is through the modification of the RadioactiveDecay data files. If in case of $^9$Li this is straightforward, whereas for $^8$He this requires some additional work since the triton decay process is not implemented in the current release. We wrote the needed class based on the alpha decay process and included it in a customized version of GEANT4.\\
As far as the radioactive decay data are concerned, with respect to the latest release (i.e. RadioactiveDecay5.3) the modification needed are the addition or modification of four files: {\it z4.a9} to describe  the $^9$Be excited levels and their decays, {\it z4.a8} to add the 11.35 MeV excited state of $^8$Be, {\it z2.a8} to modify the excited levels and Q values of $^8$He, and {\it z3.a8} to describe the triton and neutron decay of $^8$Li.
The four files can be found in appendix. Note that for the newly added states the half-life was set arbitrarily small (i.e. order of 10$^{-22}$ seconds) to have almost instantaneous decays.\\
A limit of such an implementation is that the errors on the branching ratios and on the energy of the excited states are not accounted for. To take that into account would require much deeper changes in the simulation toolkit i.e. a brand new treatment of the radioactive decays, and it would result in a second order correction as discussed in Sec.~\ref{sec:spectrum}.\\

\begin{figure} [tp]
\begin{center}
\includegraphics[width=0.75\textwidth]{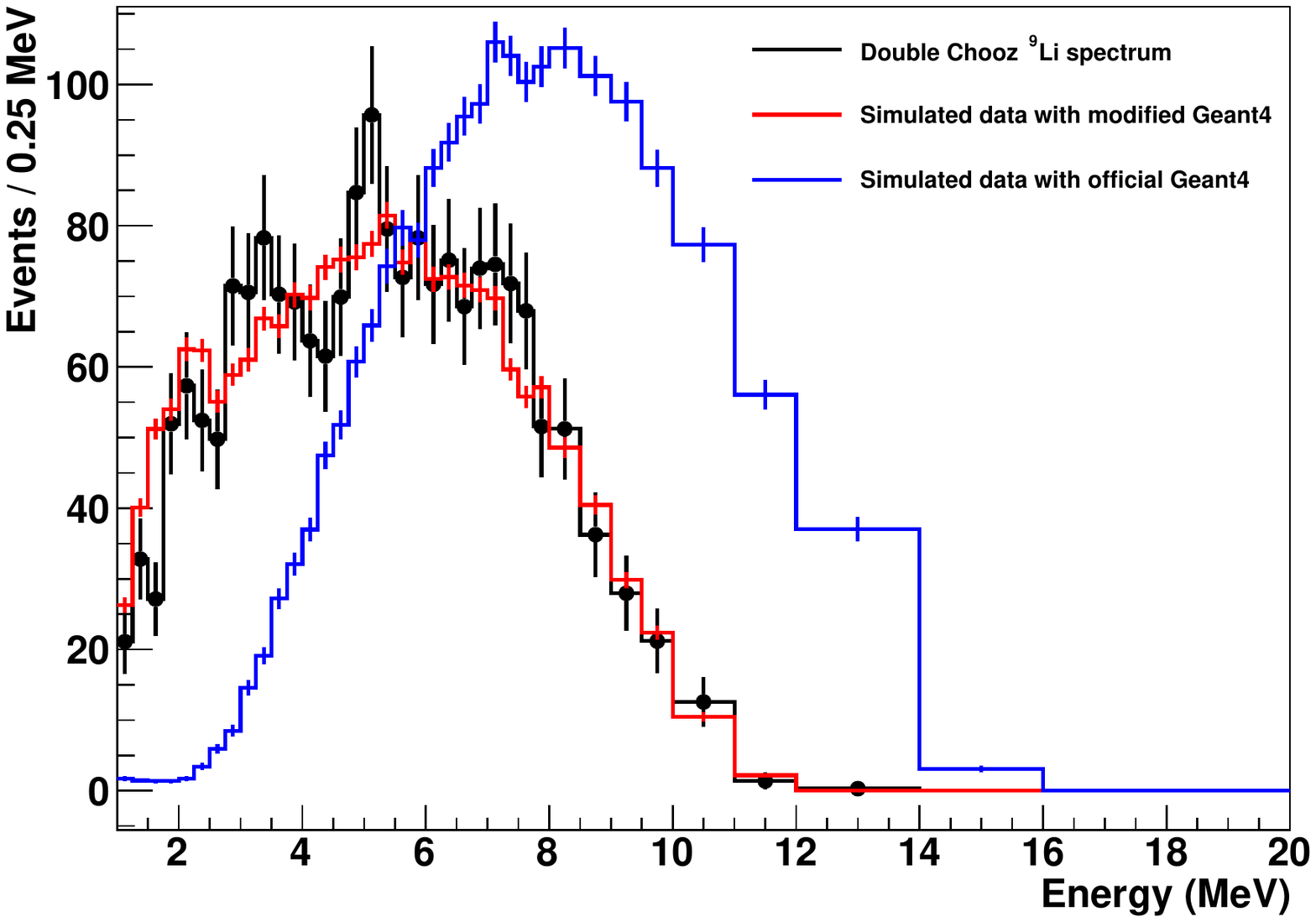}
\caption{{\it Comparison between the energy distribution of $^9$Li decays obtained with GEANT4 using the new customized radioactive decay for channels including a neutron emission (red line), GEANT4 using the standard radioactive decay for the same excited states (blue line) and the published Double Chooz distribution (black line).}}
\label{fig:3}
\end{center}
\end{figure}

\begin{figure} [t]
\begin{center}
\includegraphics[width=0.75\textwidth]{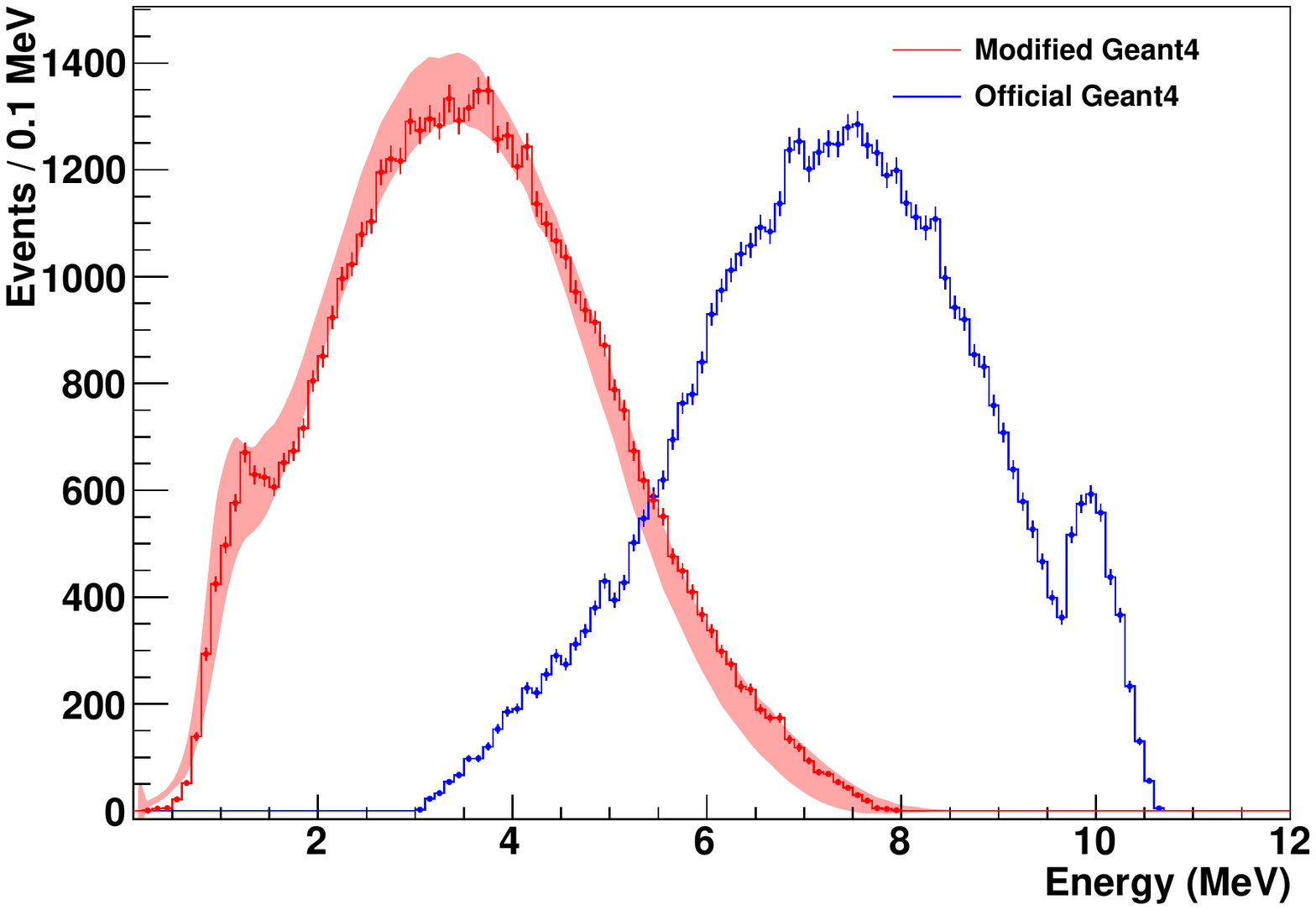}
\caption{{\it Comparison between the energy distribution  of $^8$He decays obtained with GEANT4 using the new customized radioactive decay for channels including a neutron emission (red line), and GEANT4 using the standard radioactive decay for the same excited states (blue line). The red band represents one sigma error on the shape due to the present uncertainty on energy levels and branching ratios.}}
\label{fig:4}
\end{center}
\end{figure}

With the official GEANT4 release we generated $^9$Li and $^8$He decays in a LAB (Linear Alkyl-Benzene) based liquid scintillator. Note that we included only the channels which could result in background for antineutrino reactor experiments neglecting channels without an electron and a neutron as products of the full decay chain.
We simulated a small spherical detector (5~m diameter) including scintillation (i.e. simulation of optical photons), and we applied quenching factors for $\alpha$'s, protons and tritons according to Ref.~\cite{Du:2018kyq}. A linear energy calibration to convert observed photoelectrons to visible energy was applied using as anchor point the 2.2 MeV peak given by neutron capture on Hydrogen. \\
We performed a comparison between the visible energy distributions obtained using the standard RadioactiveDecay5.3 data and the customized data package RadioactiveDecay. In addition, in case of the obtained  $^9$Li energy distribution, we benchmarked it against  the one published by the Double Chooz collaboration~\cite{DoubleChooz:2019qbj}.  The results for $^9$Li  can be found in Fig.~\ref{fig:3}. The current available radioactive decay data is clearly not suitable for a correct description of the $^9$Li decay: the absence of $\alpha$'s and neutrons replaced by gammas results in a larger energy deposition since the large quenching of $\alpha$'s is not accounted for. With the use of the customized radioactive decay data we obtained instead a very good agreement with the Double Chooz distribution: a $\chi^2$ test between the two histograms yield a $\chi^2/$n.d.f. of 1.3. However, since we have bins with low statistics a better result should be achieved with the Kolmogorov test: applied on the Monte-Carlo distribution assuming as hypothesis the distribution measured by Double Chooz, it gives a probability of 7\% of obtaining such a distribution or a worse one. \\
The same comparison was carried out on the distributions obtained for $^8$He, however in this case a benchmark on real data is not possible since Double Chooz set an upper limit on the cosmogenic $^8$He component in the background which was compatible with zero. The comparison between the spectra obtained with the released and modified GEANT4 Monte-Carlo can be seen in Fig.~\ref{fig:4}. For the customized version of GEANT4 we included the uncertainties of the energy levels and branching ratios using a toy Monte-Carlo with 200 configurations as explained in details for the $^9$Li case in Sec.~\ref{sec:spectrum} and the red band represents the  spectral shape uncertainty at one sigma. As in the case of $^9$Li, the distribution obtained with the customized version of GEANT4 is shifted to lower energies since we have neutrons and particles with large quenching factor, reducing therefore the visible deposited energy.\\

\section{Spectral uncertainty evaluation}
\label{sec:spectrum}
\begin{figure} [tp]
\begin{center}
\includegraphics[width=0.75\textwidth]{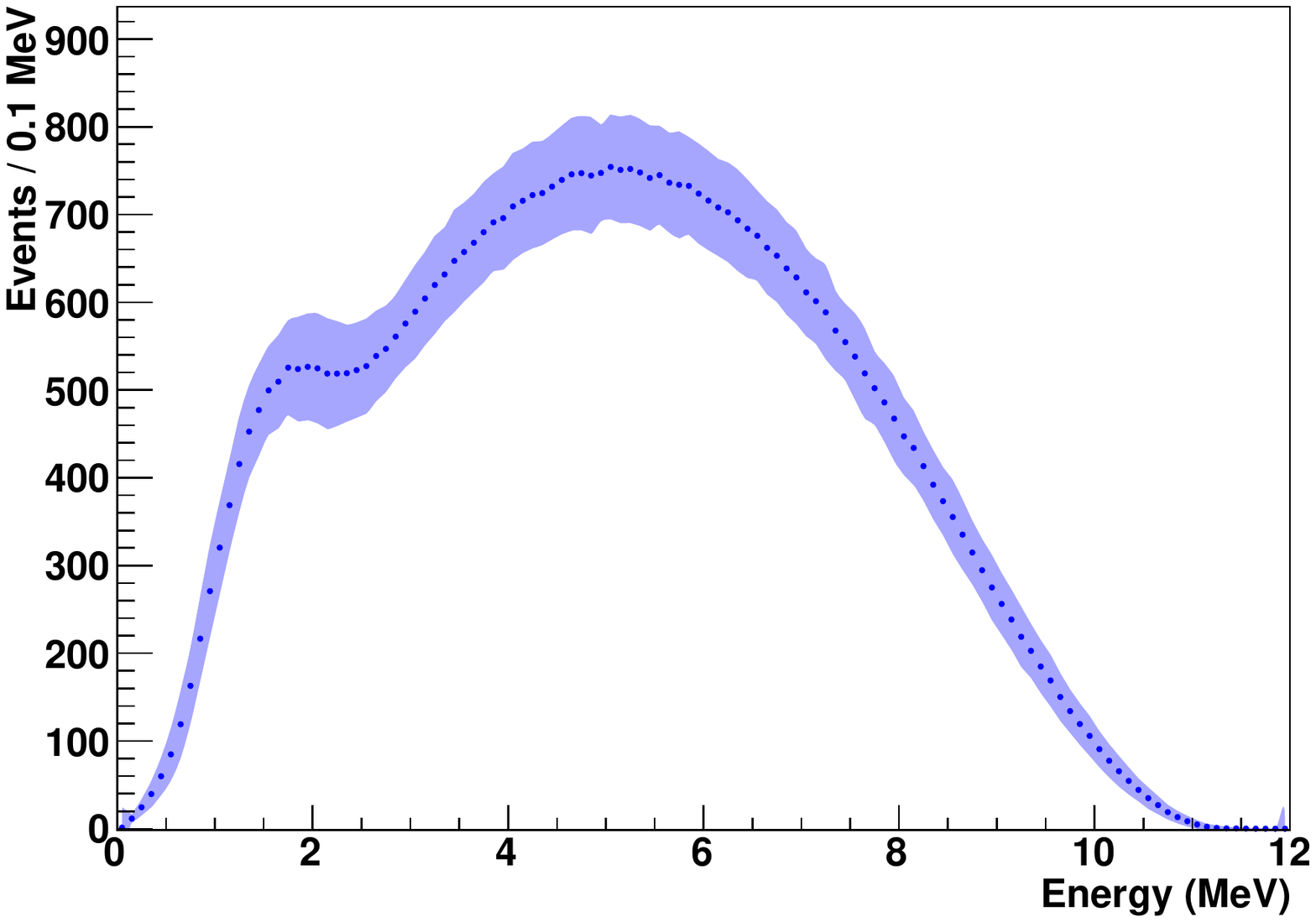}
\caption{{\it Expected visible spectrum from $^9$Li decays obtained with the new customized GEANT4. The width represents the uncertainty due to the errors on the energy levels and branching ratios.}}
\label{fig:5}
\end{center}
\end{figure}

\begin{figure} [htp]
\begin{center}
\begin{minipage}{.5\textwidth}
\includegraphics[width=0.99\textwidth]{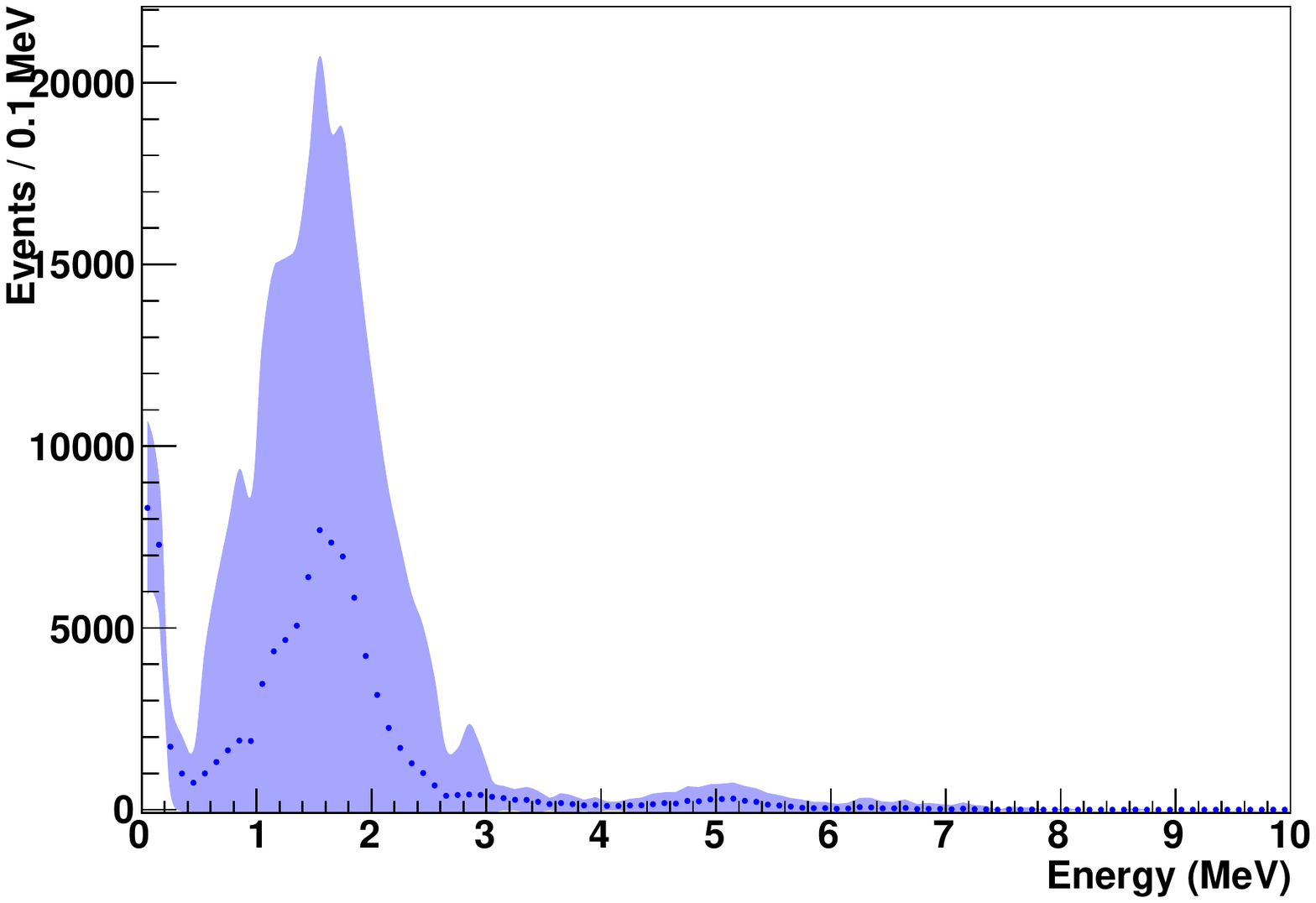}
\end{minipage}%
\begin{minipage}{.5\textwidth}
\includegraphics[width=0.99\textwidth]{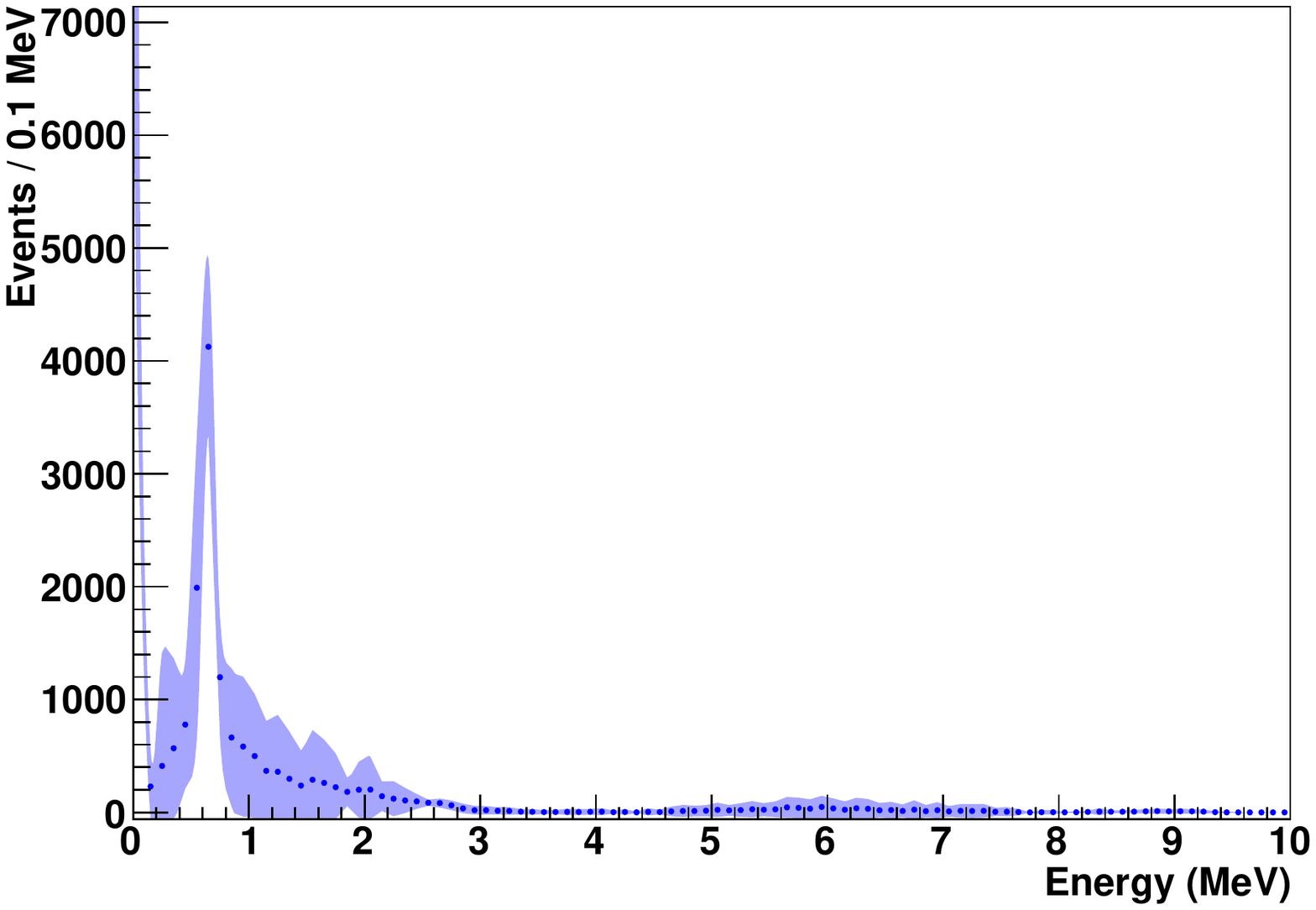}
\end{minipage}%
\caption{{\it Expected kinetic energy of $\alpha$'s (left) and neutrons (right) emitted in the $^9$Li decay chain obtained with the new customized GEANT4. The width represents the uncertainty due to the errors on the energy levels and branching ratios.}}
\label{fig:6}
\end{center}
\end{figure}

As previously mentioned, in the proposed GEANT4 implementation we do not account for the uncertainty in the energy levels and branching ratios since this would require a brand new treatment of the radioactive decays in the GEANT4 toolkit. Nonetheless we evaluated the spectral shape uncertainty with a specific toy Monte-Carlo. A set of 200 different configuration files were generated according to the known uncertainties on the branching ratios and energy levels (we assumed a Cauchy distribution) and the simulation was run in each configuration with a statistics of 50000 events.  The obtained mean visible energy with the corresponding one $\sigma$ band is shown in Fig.~\ref{fig:5}.
We clearly see that the error on the shape that we obtain generating the $^9$Li decays with the customized GEANT4 version due to the energy uncertainty of the excited levels is of second order compared to the improvement we made from the current treatment of such a decay.\\
With the same toy Monte-Carlo we estimated the kinetic energy of $\alpha$'s and neutrons as shown in Fig.~\ref{fig:6}. We did not compare them with existing measurements such as the one reported in Ref.~\cite{Nyman:1989pj}  since we should convolute them with detector dependent efficiencies, however they could be a useful input for full simulation of future experiments which could provide a cross check on these spectra.

\section{Conclusions}
Based on the available nuclear physics measurements we improved the radioactive decay treatment of the cosmogenic $^9$Li and $^8$He nuclei in GEANT4. A correct treatment of such  decays is critical for antineutrino reactor experiments since the decay products include electrons and one neutron, which mimic exactly the antineutrino signal signature. Indeed the $^9$Li background is one of the largest irreducible contribution to experiments such as Double Chooz or JUNO. The correct decay schemes are ready to be implemented in the next GEANT4 release providing an important tool for future antineutrino experiments background studies. The modified version of GEANT4 (i.e. modifications in RadioactiveDecay and new treatment of triton decay) data was tested and benchmarked against published Double Chooz data providing an excellent agreement.\\
A further improvement could be done taking into account the uncertainties on the energy levels and branching ratios. This is a second order correction and would imply a major change in GEANT4 in the way radioactive decays are treated.\\
Additional nuclear experiments studying these nuclei aiming at a better knowledge of the energy levels and branching ratios of $^9$Li, $^8$He and their decay products would be beneficial in order to further improve the simulation tools used by reactor antineutrino experiments. With a better prediction of the expected spectrum, next generation experiments such as JUNO could better constrain the production rates and in particular measure the $^8$He fraction, which is today compatible with zero due to the large uncertainties.

\newpage
\section*{Appendix}

\begin{table}[htp]
\caption{File z4.a8.}
\begin{center}
\begin{tabular}{|ccccccccc|}
\hline
\#& 8BE (8.1814e-17) &&&&&&&\\ 
\#& Excitation & Flag & Halflife & Mode & Ex & flag & Intensity & Q\\
P & 0 & - & 8.181436e-17 &&&&&\\
 &  &  & & Alpha&0& &1.&\\
 &  &  & & Alpha&0& - &100.&91.84\\
 P & 3030 & - & 1.3e-22 &&&&&\\
 &  &  & & Alpha&0& &1.&\\
 &  &  & & Alpha&0& - &100.&3121.84\\ 
P & 11350 & - & 1.3e-22 &&&&&\\
 &  &  & & Alpha&0& &1.&\\
 &  &  & & Alpha&0& - &100.&11441.84\\ 
 P & 16626 & - & 4.22e-21 &&&&&\\
 &  &  & & Alpha&0& &1.&\\
 &  &  & & Alpha&0& - &100.&16717.84\\ 
\hline
\end{tabular}
\end{center}
\label{tabAP:2}
\end{table}%

\begin{table}[htp]
\caption{File z2.a8.}
\begin{center}
\begin{tabular}{|ccccccccc|}
\hline
\#& 8HE (119.1 MS) &&&&&&&\\ 
\#& Excitation & Flag & Halflife & Mode & Ex & flag & Intensity & Q\\
P & 0 & - & 0.1191 &&&&&\\
 &  &  & & BetaMinus&0& &1.&\\
 &  &  & & BetaMinus&980.8& - &83.1&9671.0\\
 &  &  & & BetaMinus&3210& - &8.0&7441.0\\
 &  &  & & BetaMinus&5400& - &8.0&5251.0\\
 &  &  & & BetaMinus&9670& - &0.9&981.0\\
\hline
\end{tabular}
\end{center}
\label{tabAP:3}
\end{table}%

\begin{table}[htp]
\caption{File z3.a8.}
\begin{center}
\begin{tabular}{|ccccccccc|}
\hline
\#& 8LI (839.9 MS) &&&&&&&\\ 
\#& Excitation & Flag & Halflife & Mode & Ex & flag & Intensity & Q\\
P & 0 & - & 0.8399 &&&&&\\
 &  &  & & BetaMinus&0& &1.&\\
 &  &  & & BetaMinus&3030& - &100&12974.13\\
   P & 3210 & - & 4.22e-21 &&&&&\\
  &  &  & & Neutron&0& &1&\\
   &  &  & & Neutron&0& - &50&1177.7\\
 &  &  & & Neutron&477.6& - &50&700.1\\
  P & 5400 & - & 4.22e-21 &&&&&\\
  &  &  & & Neutron&0& &1&\\
   &  &  & & Neutron&0& - &50&3367.7\\
 &  &  & & Neutron&477.6& - &50&2890.1\\
 P & 9670 & - & 4.22e-21 &&&&&\\
 &  &  & & Triton&0& &0.8&\\
 &  &  & & Triton&0& - &80&4280.0\\ 
  &  &  & & Neutron&0& &0.2&\\
   &  &  & & Neutron&0& - &10&7637.7\\
 &  &  & & Neutron&477.6& - &10&7160.1\\
\hline
\end{tabular}
\end{center}
\label{tabAP:4}
\end{table}%

\begin{table}[htp]
\caption{File z4.a9.}
\begin{center}
\begin{tabular}{|ccccccccc|}
\hline
\#& 9BE (8.1814e-17) &&&&&&&\\ 
\#& Excitation & Flag & Halflife & Mode & Ex & flag & Intensity & Q\\
P & 11810 & - & 4.22e-21 &&&&&\\
 &  &  & & Alpha&0& &0.75&\\
 &  &  & & Alpha&0& - &28.&9340\\
 &  &  & & Alpha&0& - &47.&8070\\
 &  &  & & Neutron&0& &0.25&\\
 &  &  & & Neutron&0& - &2.&10140\\
 &  &  & & Neutron&3030& - &11.&7110\\
 &  &  & & Neutron&11350& - &12.&10\\
P & 11282 & - & 4.22e-21 &&&&&\\
 &  &  & & Alpha&0& &0.76&\\
 &  &  & & Alpha&0& - &76.&8812\\
  &  &  & & Neutron&0& &0.24&\\
 &  &  & & Neutron&0& - &3.&9612\\
 &  &  & & Neutron&3030& - &21.&6582\\
 P & 7940 & - & 4.22e-21 &&&&&\\
 &  &  & & Alpha&0& &0.8&\\
 &  &  & & Alpha&0& - &80.&5470\\
  &  &  & & Neutron&0& &0.2&\\
 &  &  & & Neutron&0& - &10.&6270\\
 &  &  & & Neutron&3030& - &10.&3240\\
  P & 2780 & - & 4.22e-21 &&&&&\\
 &  &  & & Alpha&0& &0.25&\\
 &  &  & & Alpha&0& - &25.&310\\
  &  &  & & Neutron&0& &0.75&\\
 &  &  & & Neutron&0& - &15.&1110\\
 &  &  & & Neutron&3030& - &60.&10\\
   P & 2429.4 & - & 4.22e-21 &&&&&\\
 &  &  & & Alpha&0& &0.025&\\
 &  &  & & Alpha&0& - &2.5&10\\
  &  &  & & Neutron&0& &0.975&\\
 &  &  & & Neutron&0& - &11.&759.4\\
 &  &  & & Neutron&3030& - &86.5&10\\
\hline
\end{tabular}
\end{center}
\label{tabAP:1}
\end{table}%


\begin{thebibliography}{99}

\bibitem{Abe:2011fz} 
  Y.~Abe {\it et al.} [Double Chooz Collaboration],
  Phys.\ Rev.\ Lett.\  {\bf 108}, 131801 (2012).
 
\bibitem{An:2015jdp} 
  F.~An {\it et al.} [JUNO Collaboration],
  J.\ Phys.\ G {\bf 43}, no. 3, 030401 (2016).
  
\bibitem{Agostinelli:2002hh} 
  S.~Agostinelli {\it et al.} [GEANT4 Collaboration],
  Nucl.\ Instrum.\ Meth.\ A {\bf 506}, 250 (2003).
  
  \bibitem{RD} 
  https://geant4.web.cern.ch/support/download
  
\bibitem{deKerret:2018fqd} 
  H.~de Kerret {\it et al.} [Double Chooz Collaboration],
  JHEP {\bf 1811}, 053 (2018).
  
\bibitem{Prezado:2005odo} 
  Y.~Prezado {\it et al.},
  Phys.\ Lett.\ B {\bf 618}, 43 (2005).

\bibitem{Cocke:1968jsh} 
  C.~L.~Cocke and P.~R.~Christensen,
  Nucl.\ Phys.\ A {\bf 111}, 623 (1968).
  
\bibitem{Tilley:2004zz} 
  D.~R.~Tilley, J.~H.~Kelley, J.~L.~Godwin, D.~J.~Millener, J.~E.~Purcell, C.~G.~Sheu and H.~R.~Weller,
  Nucl.\ Phys.\ A {\bf 745}, 155 (2004).
  
\bibitem{Prezado:2003mqr} 
  Y.~Prezado {\it et al.} [ISOLDE Collaboration],
  Phys.\ Lett.\ B {\bf 576}, 55 (2003).
  

\bibitem{Tilley:2002vg} 
  D.~R.~Tilley, C.~M.~Cheves, J.~L.~Godwin, G.~M.~Hale, H.~M.~Hofmann, J.~H.~Kelley, C.~G.~Sheu and H.~R.~Weller,
  Nucl.\ Phys.\ A {\bf 708}, 3 (2002).
  
    
\bibitem{Brown:2007zze} 
  T.~A.~D.~Brown {\it et al.},
  Phys.\ Rev.\ C {\bf 76}, 054605 (2007).
  
  
\bibitem{Papka:2007sa} 
  P.~Papka {\it et al.},
  Phys.\ Rev.\ C {\bf 75}, 045803 (2007).

\bibitem{Barker:1988biv} 
  F.~C.~Barker and E.~K.~Warburton,
  Nucl.\ Phys.\ A {\bf 487}, 269 (1988).
  
\bibitem{Borge:1993st} 
  M.~J.~G.~Borge {\it et al.} [ISOLDE Collaboration],
  Nucl.\ Phys.\ A {\bf 560}, 664 (1993).


\bibitem{Du:2018kyq} 
  Q.~Du {\it et al.},
  JINST {\bf 13}, no. 04, P04001 (2018).

  
\bibitem{DoubleChooz:2019qbj} 
  H.~de Kerret {\it et al.} [Double Chooz Collaboration],
  arXiv:1901.09445 [hep-ex].
  
\bibitem{Nyman:1989pj} 
  G.~Nyman {\it et al.} [ISOLDE Collaboration],
  Nucl.\ Phys.\ A {\bf 510}, 189 (1990).
  
\end{thebibliography}
\end{document}